\documentclass[aps,prb,twocolumn,superscriptaddress,showpacs,a4]{revtex4}
\usepackage{times}
\usepackage{amssymb,amsmath}
\usepackage{graphicx}
\usepackage{natbib}
\usepackage{multirow}
\usepackage{color}
\usepackage[a4paper,left=2cm,right=1.5cm,top=3cm,bottom=2cm]{geometry}

\newcommand{\dd}{\mathrm{d}}
\newcommand{\cplxi}{\mathrm{i}}

\begin{document}

\title{Giant microwave absorption in fine powders of superconductors}

\author{G. Cs\H{o}sz}
\affiliation{Department of Physics, Budapest University of Technology and Economics and
MTA-BME Lend\"{u}let Spintronics Research Group (PROSPIN), POBox 91, H-1521 Budapest, Hungary}

\author{B. G. M\'{a}rkus}
\affiliation{Department of Physics, Budapest University of Technology and Economics and
MTA-BME Lend\"{u}let Spintronics Research Group (PROSPIN), POBox 91, H-1521 Budapest, Hungary}

\author{A. J\'{a}nossy}
\affiliation{Department of Physics, Budapest University of Technology and Economics and
MTA-BME Lend\"{u}let Spintronics Research Group (PROSPIN), POBox 91, H-1521 Budapest, Hungary}

\author{N. M. Nemes}
\affiliation{GFMC, Unidad Asociada ICMM-CSIC "Laboratorio de Heteroestructuras con Aplicacion en Espintronica", Departamento de Fisica de Materiales Universidad Complutense de Madrid, 28040 Madrid, Spain}

\author{F. Mur\'{a}nyi}
\affiliation{Mettler-Toledo GmbH, Heuwinkelstrasse 3., CH-8606 N\"anikon, Switzerland}

\author{G. Klupp}
\affiliation{Institute for Solid State Physics and Optics, Wigner
Research Centre for Physics, Hungarian Academy of Sciences, P.O. Box
49, H-1525 Budapest, Hungary}

\author{K. Kamar\'{a}s}
\affiliation{Institute for Solid State Physics and Optics, Wigner
Research Centre for Physics, Hungarian Academy of Sciences, P.O. Box
49, H-1525 Budapest, Hungary}

\author{V. G. Kogan}
\affiliation{Ames Laboratory, U.S. Department of Energy and Department of Physics and Astronomy, Iowa State University,
Ames, Iowa 50011, USA}

\author{S. L. Bud'ko}
\affiliation{Ames Laboratory, U.S. Department of Energy and Department of Physics and Astronomy, Iowa State University,
Ames, Iowa 50011, USA}

\author{P. C. Canfield}
\affiliation{Ames Laboratory, U.S. Department of Energy and Department of Physics and Astronomy, Iowa State University,
Ames, Iowa 50011, USA}

\author{F.\ Simon}
\email{f.simon@eik.bme.hu}
\affiliation{Department of Physics, Budapest University of Technology and Economics and
MTA-BME Lend\"{u}let Spintronics Research Group (PROSPIN), POBox 91, H-1521 Budapest, Hungary}

\begin{abstract}

Enhanced microwave absorption, larger than that in the normal state, is observed in fine grains of type-II superconductors (MgB$_2$ and K$_3$C$_{60}$) for magnetic fields as small as a few $\%$ of the upper critical field. The effect is predicted by the theory of vortex motion in type-II superconductors, however its direct observation has been elusive due to skin-depth limitations; conventional microwave absorption studies employ larger samples where the microwave magnetic field exclusion significantly lowers the absorption. We show that the enhancement is observable in grains smaller than the penetration depth. A quantitative analysis on K$_3$C$_{60}$ in the framework of the Coffey--Clem (CC) theory explains well the temperature dependence of the microwave absorption and also allows to determine the vortex pinning force constant.
\end{abstract}
\maketitle

\section{Introduction} 
Electrodynamics of superconductors remains an intensively studied field \cite{DresselReview} due to the wealth of attainable fundamental information, including the nature of pairing mechanism and the coupling strength, and also due to the technological importance of these materials. As an example, observation of the conductivity coherence peak in conventional superconductors\cite{HolczerPRB} (Nb and Pb) and its absence in high-$T_{\text{c}}$ materials \cite{YBCO_no_coh_peak} pointed to a BCS mechanism in the former and it was an early indication of non-BCS superconductivity in the latter compounds. Concerning applications, the DC electrodynamic properties in the mixed state of type-II superconductors determine the utility (e.g. loss, permanent field homogeneity, and stability) in superconducting solenoid coils that are widely used in superconducting particle acceleration, solid state spectroscopy, or medical imaging. The AC electrodynamic properties are relevant for applications including e.g. power handling, sound and electromagnetic field detection \cite{SinglePhotonSupercDet,SupercPhotonDetGraphene,THzSuperconductor}, superconducting microwave resonators \cite{LancasterBook}, and in microwave absorbers \cite{FronciszNature}.

The frequency dependent conductivity of superconductors, $\widetilde{\sigma}=\sigma_1+\cplxi \sigma_2$, is well known for both BCS (i.e. weak-coupled \textit{s}-wave pairing) and non-BCS superconductors (including strongly coupled \textit{s}-wave and non \textit{s}-wave superconductors) in the absence of magnetic field, $B=0$. At zero temperature, $T=0$, the real part, $\sigma_1(\omega)$, is a delta function at $\omega=0$ followed by $\sigma_1(\omega)=0$ until the gap edge at $\omega_{\text{g}}=2\Delta/\hbar$ (Ref. \onlinecite{TinkhamBook}) (usually at $0.1-10$ THz). According to the Ferrell--Glover--Tinkham (FGT) sum rule \cite{FerrelGloverPRB1958,TinkhamFerrellPR1959}, the spectral weight of the delta function comes from states which are gapped below $\omega_{\text{g}}$ (the sum rule is discussed in depth in the Supplementary Material). The technologically important radio frequency range spans $9$ orders of magnitude in superconductors (from $10$ kHz up to $1$ THz) with similar characteristic properties, thus measurements in the microwave range ($1-100$ GHz) are representative.

Conductivity in finite magnetic fields for the mixed state in type-II superconductors was first described by the Bardeen--Stephen (BS) model \cite{BardeenStephen} for the viscous motion of vortices. This was later improved by the Coffey--Clem theory (CC) in a series of seminal papers \cite{CC91,CC921,CC922,CC923,CC924,CC93}, which also includes the effect of pinning force on the vortex motion. The most important prediction of the BS model is a finite $\sigma_1$ conductivity at $\omega=0$. However, it is less well-known that the FGT sum rule implies a non-zero $\sigma_1$ that is \emph{larger} than in the normal state for $\omega<\omega_{\text{g}}$. Observation of this effect has been elusive as most contributions study the surface impedance on polycrystalline \cite{HolczerPRB}, compacted powder pellet, or thin film samples \cite{MgB2ThinFilm,NbThinFilms}. Surface impedance studies have the advantage that sample geometry is well defined, however the effects of $\sigma_1$ and $\sigma_2$ are inevitably intermixed in this type of measurements. Given that $\sigma_2$ is orders of magnitude larger than $\sigma_1$ in the superconducting state (due to the small value of the penetration depth, $\lambda$, with respect to sample thickness), the surface impedance measurement is less sensitive to changes in $\sigma_1$ (Refs. \onlinecite{HendersonPRL1998, RevenazPRB1994, BelkPRB1996,ParksPRL1995,WuIEEE1993}). 

The effects of $\sigma_1$ and $\sigma_2$ are decoupled for fine grains; for a sample placed in a microwave cavity, the loss is due to $\sigma_1$, whereas the resonance shift is due to $\sigma_2$ (Refs. \onlinecite{buravov71,Klein1993,MaedaPRL}). Therefore such samples provide a unique opportunity to test the predictions of the CC theory on $\sigma_{1,2}(T,B)$. The enigmatic and yet unexplained increase of the electron spin resonance signal in superconductors right below $T_{\text{c}}$ (Refs. \onlinecite{SimonPRB2005,MuranyiPRB2008}) also highlights the need to study further the electromagnetic absorption in superconductors.

This motivated us to revisit the microwave conductivity (at about $10$ GHz) in the MgB$_2$ and K$_3$C$_{60}$ superconductors as a function of $T$ and $B$. We observe an excess microwave loss (or microwave absorption) in small magnetic fields (as low as a few $\%$ of the upper critical field, $B_{\text{c2}}$) for a sample consisting of well-separated fine grains (typical size is a few microns). The excess microwave absorption is not observable in a single crystal sample. A quantitative analysis is provided for K$_3$C$_{60}$, which is a one-gap, cubic superconductor with well known magneto-transport properties \cite{GunnRMP}, whereas MgB$_2$ is a multi-band superconductor with strongly anisotropic $B_{\text{c2}}$ (Refs. \onlinecite{MgB2_Nature,Buzea}), thus application of the CC model is less straightforward. 

\section{Methods and Experimental} 
We studied fine powder MgB$_2$ samples identical to batches in Refs. \onlinecite{SimonPRL2001} and \onlinecite{SimonPRB2005b}. Single crystal and powder K$_3$C$_{60}$ samples were prepared by the conventional K intercalation method; the crystal sample was from the same batch as in Ref. \onlinecite{NemesPRB2000}. The powder samples were further ground together with non-conducting SnO$_2$ powder to prevent conducting links between the grains. Magnetometry attested that static superconducting properties (such as the steepness of the superconducting transition) were unaffected by the mixing. Samples were sealed in quartz ampules under low pressure helium. Microwave properties were measured with the cavity perturbation method \cite{buravov71,Klein1993} as a function of temperature, $T$, and in various static magnetic fields, $B$, inside a superconducting solenoid, with zero-field cooling. Zero field measurements (besides the Earth's magnetic field) were made in another cryostat without a magnet field solenoid to avoid trapped flux (which may amount to 10-20 mT). The unloaded copper cavity has a quality factor, $Q_0 \sim 10,000$ and a resonance frequency, $f_0 \sim 11.2$ GHz, whose temperature dependence is taken into account. The samples were placed in the node of the microwave (or \textit{rf}) electric field and maximum of the microwave magnetic field inside the TE011 cavity, which is the appropriate geometry to study minute changes in the conductivity \cite{MaedaPRL}. The \textit{rf} magnetic field is parallel to the DC field of the solenoid, which yields the largest vortex motion induced absorption according to the CC theory \cite{CC922}. Measurement \cite{MehringRSI} of the quality factor, $Q$, and the cavity resonance frequency, $f$ yields the loss: $\Delta\left(\frac{1}{2Q}\right)=\frac{1}{2Q}-\frac{1}{2Q_0}$ and cavity shift: $\Delta f/f_0=(f-f_0)/f_0$. 

\begin{figure}[h!]
\begin{center}
\includegraphics[width=1\linewidth]{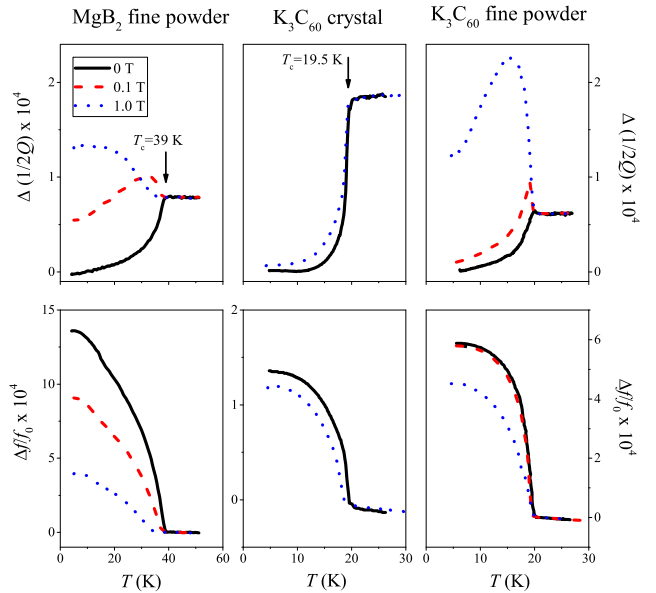}
\caption{Temperature dependent cavity loss, $\Delta(\frac{1}{2Q})$, and cavity shift, $\Delta f/f_0$ for a fine powder of MgB$_2$ and for the single crystal and powder K$_3$C$_{60}$ samples. Two magnetic field data are shown for the crystal ($0$ and $1$ T) and three for the powder samples ($0$, $0.1$, and $1$ T). Note that the cavity loss changes significantly for the powder sample in contrast to the single crystal sample. Note the different scales for the $\Delta f/f_0$ data.}
\label{fig:Fig1_expdata}
\end{center}
\end{figure}

\section{Results} 
Fig. \ref{fig:Fig1_expdata} shows the microwave cavity loss and cavity shift for a fine powder of MgB$_2$ and for two kinds of K$_3$C$_{60}$ samples: a single crystal and a fine powder as a function of temperature and for a few magnetic field values. The microwave loss decreases rapidly below $T_{\text{c}}$ in zero magnetic field as expected for superconductors. The most important observation is that the microwave loss becomes significant for a magnetic field as small as $0.1$ T for the fine powder samples, whereas even $1$ T has little effect on the microwave absorption for the single crystal K$_3$C$_{60}$ sample. In fact, we observe a \emph{giant}, about a factor $3$ times larger, microwave absorption below $T_\text{c}$ than in the normal state. This striking difference between the crystal and fine grain samples is clearly demonstrated for K$_3$C$_{60}$ where measurements on both kinds of samples are shown. For MgB$_2$, microwave measurements on compacted samples (or surface impedance measurements) supports this observation as therein no enhanced microwave absorption was observed\cite{MgB2_mwcond1,MgB2_mwcond2,MgB2_mwcond3,MgB2_mwcond4,MgB2_mwcond5,MgB2_mwcond6}. While the absorption becomes significant for the fine powder samples at $B=0.1$ T, the shift changes less, which means that the overall superconducting characteristics of the sample are maintained.

We believe that the enhanced microwave absorption is an ubiquitous property of fine powders of type-II superconductors. However, we cannot quantitatively discuss this effect for MgB$_2$ due to the multi-band superconductivity \cite{MazinPRL2001,LouieNat2002} and the significant $B_{\text{c2}}$ anisotropy; $B_{\text{c2}}$ at 0 K is $\sim2\,\text{T}$ and $\sim16\,\text{T}$ for $B||(c)$ and $B||(a,b)$, respectively (Refs. \onlinecite{SimonPRL2001,AnisotropyCanfield,Buzea}). We therefore focus on K$_3$C$_{60}$ in the following. The enhanced microwave loss appears progressively with increasing magnetic field (additional data are shown in the Supplementary Material). 

We also show the $B=0.1$ T data for the powder sample ($B\approx 0.002 \times B_{\text{c2}}$) in Fig. \ref{fig:Fig1_expdata}; they show a peak in the microwave loss right below $T_{\text{c}}$ followed by a gradual decrease. The zero magnetic field data also shows a small peak (invisible at the scale of Fig. \ref{fig:Fig1_expdata}) for the powder sample (shown in the Supplementary Material). This small peak is not due to magnetic field and is most probably a tiny conductivity coherence peak (the analogue of the Hebel--Slichter peak \cite{HebelSlichter}) which is known to be strongly suppressed by strong-coupling effects in alkali fullerides \cite{MacFarlaneMuSR,PenningtonRMP}. While the presence of a coherence peak itself is an interesting physical phenomenon\cite{HolczerPRB,DresselArxiv2018}, it is not relevant for the present discussion.

The fact that the enhanced microwave absorption occurs with the application of the magnetic field hints at a flux motion related phenomenon that is discussed in the framework of the CC theory. The microwave absorption peak occurs above the irreversibility line, i.e. it is related to the physical behavior of the vortex-fluid state; for K$_3$C$_{60}$ $T_{\text{irr}}(B=0.1\,\text{T})\approx15\,\text{K}$ and $T_{\text{irr}}(B=1\,\text{T})<5\,\text{K}$ (Ref. \onlinecite{PrassidesBook}).

The strong dependence on the sample morphology is also discussed below. Superconducting fullerides are type-II ($\lambda\gg\xi$) and have a short mean free path\cite{GunnRMP} i.e. they can be described in the local electrodynamics limit as opposed to the non-local (or Pippard) limit, which simplifies the discussion. 	
			
\section{Discussion} 
\subsection{Conductivity in the superconducting state}
The phenomenological CC theory \cite{CC91,CC921,CC922,CC923,CC924,CC93} is based on a two-fluid model and considers the motion of vortices due to the exciting electromagnetic field in the presence of a viscous background (described by the viscous drag coefficient, $\eta$) and a restoring force (described by an effective pinning force constant, $\kappa_{\text{p}}$). 

The viscous drag was introduced in the Bardeen--Stephen theory \cite{BardeenStephen} and is determined by the superconducting parameters \cite{TinkhamBook,PatrickLaurent}: $\eta\left(T\right)=\frac{\Phi_\text{0}B_\text{c2}\left(T\right)}{\rho_\text{n}\left(T\right)}$. The value of $\kappa_{\text{p}}$ is unknown and only an upper limit can be estimated from thermodynamic considerations \cite{Anderson,WuSridhar,PatrickLaurent} for a "perfect pinning center", i.e. a hollow cylinder with a diameter of about the coherence length, $d\approx \xi$. The condensation energy gain per unit length from placing a vortex in this cylinder is about $d^2 B_{\text{c}}^2/2\mu_0$, where the square of the thermodynamic critical field is $B_{\text{c}}^2=B_{\text{c1}} B_{\text{c2}}/\ln\left(\frac{\lambda}{\xi}\right)$. This leads to: 
\begin{equation}
\kappa_{\text{p,max}}=\frac{B_\text{c}^2}{2\mu_0}.
\label{eq:kappa_p,max}
\end{equation} 
For a weaker pinning center, $\kappa_{\text{p}}$ can be significantly lower and in the bulk of a perfect superconductor, it would be zero.

The CC theory introduces the concept of the complex penetration depth, $\widetilde{\lambda}$:
\begin{equation}
	\widetilde{\lambda}^2=\frac{\lambda^2+(\cplxi/2)\widetilde{\delta}_\text{vc}^2}{1-2 \cplxi\lambda^2/\widetilde{\delta}_\text{nf}^2},
	\label{eq:CC_lambda}
\end{equation}
where $\widetilde{\delta}_\text{nf}$ is the normal fluid skin depth, $\lambda$ is the usual (real) penetration depth and $\widetilde{\delta}_\text{vc}$ is the complex effective skin depth \cite{CC93}. The latter quantity is zero for $B=0$ and becomes finite in the mixed state when vortex motion is present. $\widetilde{\lambda}$ is related to the complex conductivity by $\widetilde{\sigma}=\cplxi/\mu_{\text{0}}\omega\widetilde{\lambda}^2$. Note that for $B=0$ (i.e. when $\widetilde{\delta}_\text{vc}^2=0$), we obtain $\widetilde{\lambda}^2(T=0)=\lambda^2$ and $\widetilde{\lambda}^2(T=T_{\text{c}})=\cplxi \delta_{\text{n}}^2/2$ as expected. The CC theory yields the temperature and magnetic field dependent $\widetilde{\sigma}$ using explicit expressions for $\lambda(T)$, $\delta_\text{nf}(T)$, $B_\text{c2}(T)$.  We have implemented the calculation (details are given in the Supplementary Material) and validated our calculations by comparing the results to that published in Ref. \onlinecite{CC93}. 

\begin{figure}[h!]
	\begin{center}
	\centering
    \includegraphics[width=1\linewidth]{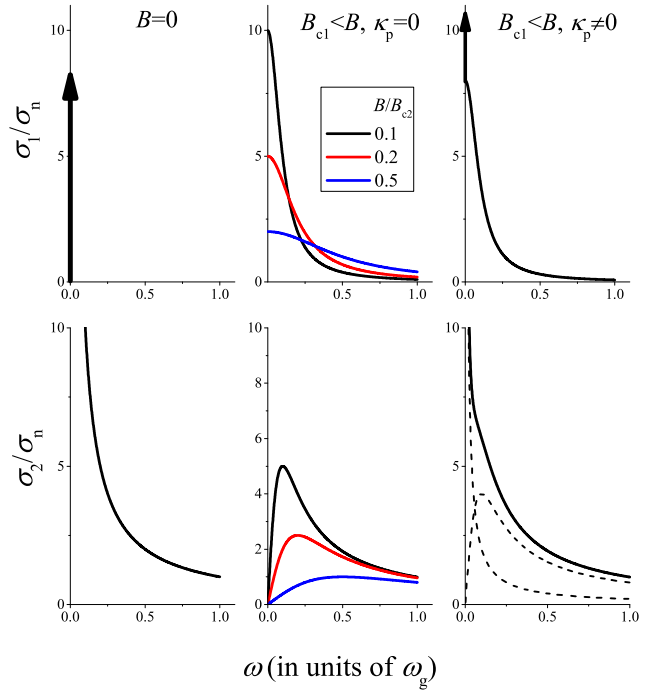}
	\caption{Illustration of $\widetilde{\sigma}(\omega)$ in superconductors for i) $B=0$, ii) for finite fields ($B>B_{\text{c1}}$) with $\kappa_{\text{p}}=0$ (the Bardeen--Stephen case), and iii) for $B\neq0$ and a finite $\kappa_{\text{p}}$ (the case of the CC theory). Conductivity above the superconducting gap, $\omega_{\text{g}}$, is not shown. The spectral weight in the delta function is preserved for $B\neq 0$. Note that for $\kappa_{\text{p}}\neq 0$, the conductivity appears as if it were a sum of $\widetilde{\sigma}$'s for the $B=0$ and the BS flux-flow regimes (shown with dashed curves). Of the two components, the $\sigma_1\propto \delta(\omega)$ and $\sigma_2\propto 1/\omega$ is due to vortex pinning.}
	\label{fig:Fig2_sigma_vs_freq_DEMO}
	\end{center}
\end{figure}
	
Here, we discuss qualitatively the predictions of the CC theory and some typical cases are shown in Fig. \ref{fig:Fig2_sigma_vs_freq_DEMO}. In superconductors, at $B=0$ the carrier spectral weight below $\omega_\text{g}$ collapses into the $\sigma_1 =\frac{\pi}{2\mu_0\lambda^2} \delta(\omega)$ function according to the Ferrell--Glover--Tinkham (FGT) sum rule \cite{FerrelGloverPRB1958,TinkhamFerrellPR1959}. The Kramers--Kronig relation dictates that $\sigma_2 =1/\mu_0\omega \lambda^2$. Without vortex pinning, the Meissner state is destroyed for $B>B_{\text{c1}}$ and the Bardeen--Stephen theory gives $\sigma=\frac{B_{\text{c2}}}{B}\frac{\sigma_{\text{n}}}{1+\cplxi\frac{\omega}{\omega_{\text{c}}}}$. It is worth noting that this result is formally analogous to the AC Drude model as the underlying equation of motion (of electrons or vortices) is the same. Here, we introduced a cut-off frequency $\omega_{\text{c}}=\frac{B}{\mu_0 \lambda^2 B_\text{c2}\sigma_\text{n}}$. Clearly, $\sigma_1$ can be larger than $\sigma_\text{n}$ for $\omega<\omega_\text{c}$.

In the presence of vortex pinning, the CC theory predicts that $\widetilde{\sigma}$ is characterized by a mixture of the unperturbed superconducting behavior and that of the Bardeen--Stephen theory with a shared spectral weight which depends on the pinning force constant. Pinning reduces the effect of the vortex flow on $\sigma_1$. The enhanced $\sigma_1(\omega)$ AC conductivity (as compared to the normal state) is a direct consequence of the FGT sum rule for a finite magnetic field. It allows to estimate the maximum possible value of the enhancement as $\sigma_{1,\text{max}}(\omega)\approx\sigma_{\text{n}} \times\omega_{\text{g}}/\omega$, that would be realized at $T=0$ in the absence of pinning. E.g. for K$_3$C$_{60}$ and $\omega/2\pi=10\,\text{GHz}$ we obtain $\sigma_{1,\text{max}}(\omega)\approx 140\sigma_{\text{n}}$.

The CC theory allows to quantitatively analyze the conductivity in K$_3$C$_{60}$. The requirement of $B\gg B_{\text{c1}}$ is satisfied for our magnetic fields of $0.1 \dots 1$ T as $B_{\text{c1}}\sim 10$ mT. The CC theory was developed for a superconductor which occupies the total half space. We show in the Supplementary Material that it can be applied for a spherical sample which approximates well finite sized grains containing at least a few hundred/thousand vortices. In addition, the static and \textit{rf} magnetic fields are parallel in our experiment, which is the standard case for the applicability of the CC theory. Albeit we cannot quantitatively consider the effect of the small particle size on the magnetic properties, we believe that neither the surface barriers (also known as Bean-Livingstone barriers\cite{BeanLivingstone}) nor the so-called geometrical barriers \cite{ZeldovPRL1994} affect considerably the applicability of the CC theory. The argument is that both types of barriers would affect the overall number of the vortices under the applied DC magnetic field (or the $B$ value where vortices appear) but not the overall vortex dynamics under the application of the small AC magnetic field, which is the primary reason for the observed microwave absorption.

\begin{table}[h!]
	\centering
\begin{tabular}{l c r}
\hline \hline
property & value & Ref.\\
\hline
	$T_\text{c}$ & $19.5$ K & \onlinecite{GunnRMP} \\
	$\rho_{\text{n}}(T_{\text{c}})$ &  $1.8\cdot 10^{-6},\,4.1\cdot 10^{-6}$ $\Omega$m & \onlinecite{ResDBSF,ResKlein}\\
	$\delta_\text{n}(11.1\,\text{GHz})$ & $9.7,\,6.4$ $\mu$m & \\
	$\xi_0$ & $2.6,\,3.4$ nm & \onlinecite{Critfield,Upcritfield} \\
	$\lambda_0$ & $240,\,480,\,600$ nm & \onlinecite{Critfield,UemuraNat91,PendepthNMR} \\
	\hline
\end{tabular}
\caption{Transport and magnetic parameters of the K$_3$C$_{60}$ superconductor: the superconducting transition temperature, $T_\text{c}$; the normal state resistivity at $T_{\text{c}}$, $\rho_{\text{n}}$; the normal state skin depth, $\delta_\text{n}$; the coherence length at $T=0$, $\xi_0$; and the magnetic field penetration depth at $T=0$, $\lambda_0$. The tabulated $\xi_0$ values correspond to an upper critical field, $B_{\text{c2}}$ at $T=0$ of 49 and 28 T, respectively.}
\label{tab:prop}
\end{table}

\begin{figure}[h!]
	\begin{center}
	\centering
   \includegraphics[width=1\linewidth]{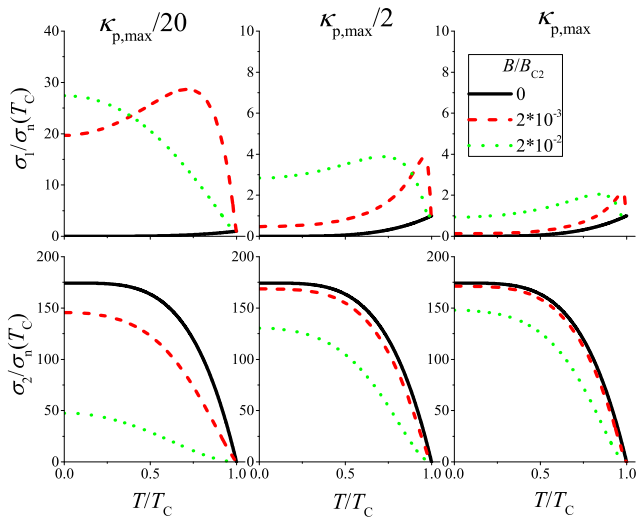}
	\caption{Calculated real part and imaginary part of complex \textit{rf} conductivity vs the reduced temperature for different values of the pinning force constant, $\kappa_{\text{p}}$. The conductivity values are normalized by the normal state conductivity at the critical temperature. The large value of $\sigma_2(T=0)/\sigma_{\text{n}}(T_{\text{c}})$ is due to a large $\left(\delta_{\text{n}}/\lambda\right)^2$. Note also the different scales for the $\sigma_1$ values.}
	\label{fig:Fig3_sigma}
	\end{center}
\end{figure}
In Fig. \ref{fig:Fig3_sigma}., the calculated conductivity is plotted versus the reduced temperature for different force constants, $\kappa_{\text{p}}$, with the parameters $\xi_0=3\,\text{nm}$ (which corresponds to $B_{\text{c2}}(\text{0 K})=37.5$ T), $\lambda_0=440$ nm, and $\rho_{\text{n}}(T_{\text{c}})=2.95$ $\Omega$m, the mean values of the corresponding literature parameters for K$_3$C$_{60}$ (Refs. \onlinecite{ResDBSF,ResKlein,Critfield,Upcritfield,UemuraNat91,PendepthNMR,GunnRMP}), which are detailed in Table \ref{tab:prop}. The pinning force constant is given in the figure with respect to $\kappa_{\text{p,max}}=3.84\cdot 10^4$ $\text{N}/\text{m}^2$ according to Eq. \eqref{eq:kappa_p,max}. The figure indicates that $\kappa_{\text{p}}$ strongly affects the magnetic field dependence of $\widetilde{\sigma}$.

\subsection{Analysis of the experimental data}
The sample morphology greatly affects the relation between the material conductivity, $\widetilde{\sigma}$, and the microwave parameters, the loss and shift. Two limiting cases are known. 1) The sample is large and the field penetrates into a limited distance from the surface. This approximates the measurement in the large K$_3$C$_{60}$ single crystal. 2) The sample is a small sphere with radius comparable to the penetration depth. This approximates the K$_3$C$_{60}$ sample of well divided small grains. We discuss that the experimental observations for the crystal and fine powder K$_3$C$_{60}$ are explained well by these two regimes. 

In the first case, when the \textit{rf} field penetrates in the skin depth only (known as the skin limit), the following equation holds between the microwave measurement parameters and the material quantities \cite{Klein1993}:
\begin{gather}
	\frac{\Delta f}{f_\text{0}}-\cplxi\Delta\left(\frac{1}{2Q}\right)=-\cplxi \nu \mu_0 \omega \sqrt{-\widetilde{\lambda}^2}, 
\label{eq:skin_lim}   
\end{gather}
where the complex penetration depth, $\widetilde{\lambda}$, is related to the conductivity as $\widetilde{\lambda}^2=\cplxi(\mu_{\text{0}}\omega\widetilde{\sigma})^{-1}$. The dimensionless $\nu\ll 1$ is the so-called resonator constant \cite{HolczerPRB} and it depends on the sample surface relative to that of the cavity. 

\begin{figure}[h!]
	\begin{center}
		\centering
		\includegraphics[width=1\linewidth]{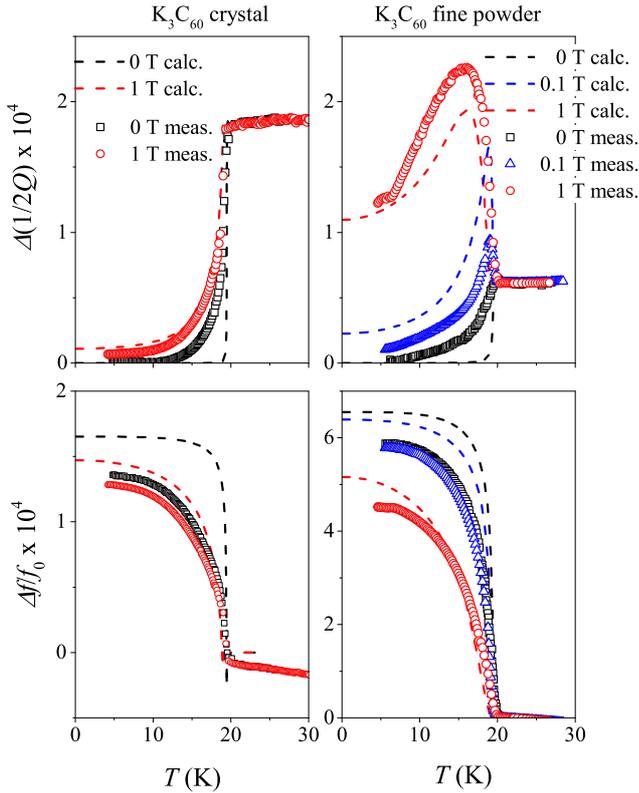}
		\caption{Comparison of measured and calculated cavity loss and shift parameters in the skin (left panels) and penetration limit (right panels). Calculation details are given in the text. Note that the calculated curves and the experimental data agree well for both sample types.}
		\label{fig:Fig4_meas_sim_comp}
	\end{center}
\end{figure}
The left panels in Fig. \ref{fig:Fig4_meas_sim_comp}. show the calculated and measured cavity loss and shift in 0 and 1 T magnetic fields for the single crystal sample. The calculation uses Eq. (3) with $\kappa_{\text{p}}=\kappa_{\text{p,max}}/20$. Although this low $\kappa_{\text{p}}$ induces a large $\sigma_1$, there is no visible peak in the cavity loss in this case when excitation is limited to the surface. We discuss in detail in the Supplementary Material that the calculated cavity loss and shift are insensitive to the value of $\kappa_{\text{p}}$ in this limit. Clearly, the experimental data for the K$_3$C$_{60}$ crystal match well the calculations. 

A suitable $\nu=5.1\cdot 10^{-4}$ was chosen to match the calculation to the experiment. We find that for both the calculation and experiment, the cavity loss parameter drops rapidly below $T_{\text{c}}$, although $\sigma_1/\sigma_{\text{n}}$ is around unity due to the vortex motion. This effect is due to the development of a significant $\sigma_2/\sigma_{\text{n}}\sim 100$, which limits the penetration of microwaves into the sample and thus reduces the loss. This means that the microwave surface impedance measurement is not capable of providing information about $\sigma_1$ in the presence of vortex motion. We note that the experimental curves do not show such a rapid change as a function of temperature as the calculation. This may be related to the finite size and surface roughness of the single crystal sample.

Second, we discuss the opposite limit, when the microwave field penetrates into the sample (known as the penetration limit), the cavity measurables depend differently on the sample parameters. It was shown\cite{LandauBook} for a sphere with radius, $a$:
\begin{gather}
	\frac{\Delta f}{f_\text{0}}-\cplxi\Delta\left(\frac{1}{2Q}\right)=- \gamma \widetilde{\alpha}, \\
	\widetilde{\alpha}=-\frac{3}{2}\left(1-\frac{3}{a^2 \widetilde{k}^2}+\frac{3}{a \widetilde{k}}\cot(a \widetilde{k}) \right),
\label{eq:penetr_lim}   
\end{gather}
where $\widetilde{k}=\widetilde{n}\frac{\omega}{c}$ is the complex wavenumber, with $\widetilde{n}=\sqrt{\cplxi\widetilde{\sigma}/\epsilon_\text{0}\omega}$ being the complex index of refraction. The dimensionless $\gamma$ is a sample volume dependent constant.

The right panels in Fig. \ref{fig:Fig4_meas_sim_comp}. show the measured cavity loss and shift data for the fine powder sample together with a fit according to Eq. \eqref{eq:penetr_lim}. To obtain these fits, we fixed the transport and magnetic parameters ($\rho_{\text{n}}$, $\xi_0$, $\lambda_0$) of K$_3$C$_{60}$ to the respective mean values as given in Table \ref{tab:prop}. We assumed that the sample consists of spheres with a uniform diameter, $a$. The zero magnetic field data depends only on $\gamma$ and $a$ when the other parameters, $\delta_{\text{n}}$ and $\lambda$, are fixed. A fit to the $B=0$ data yields $\gamma=5.5(2)\cdot10^{-4}$, and $a=6.2(2)$ $\mu$m. We then proceed to fit the magnetic field dependent data with $\kappa_{\text{p}}$ as the only free parameter and we obtain $\kappa_{\text{p}}=1.0(1) \cdot 10^3$ $\text{N}/\text{m}^2$, which is about $\kappa_{\text{p,max}}/20$. As shown in Fig. \ref{fig:Fig4_meas_sim_comp}., the calculation agrees well with the experimental data.

Somewhat better fits could be obtained when letting $\lambda$, $\rho_{\text{n}}$, and $B_{\text{c2}}$ differ from the mean literature values. In addition, Eq. \eqref{eq:penetr_lim} is valid for spheres only, it thus fixes the ratio between the real and imaginary parts (cavity loss and shift). A different particle shape or particle size distribution would allow for a different scaling factor for the loss and shift data which could also improve the fits. Although improved fits could be attained, we believe that the simplest model explains well the experimental observation of an enhanced microwave absorption. In addition it allows to determine an effective pinning force constant, which is an important parameter to describe the electrodynamics of type-II superconductors. However, we note that $\kappa_{\text{p}}$ determined herein may overestimate the bulk pinning force constant; it is known that the presence of a substantial surface-volume ratio may give rise to additional vortex pinning\cite{BeanLivingstone,ZeldovPRL1994}, with a strength that is difficult to estimate.

\section*{Summary} 
We demonstrated that moderate magnetic fields, which are small compared to the upper critical field, induces a large microwave absorption in fine powders of type II superconductors, like MgB$_2$ and K$_3$C$_{60}$.
The effect is absent for samples containing larger grains or compacted powder pellets. The Bardeen--Stephen model of flux-flow predicts that the real part of the AC conductivity can be enhanced in the microwave range, but this effect has not been observed. We analyze the conductivity using the Coffey--Clem theory which also accounts for vortex pinning effects. It is applied to calculate the microwave properties for two kinds of samples: when the electromagnetic field penetration is limited to the surface (skin limit) or when it fully penetrates into the fine grain samples (penetration limit). We show that microwave absorption in the skin limit is little affected by the vortex-motion enhanced $\sigma_1$ but in the penetration limit, the effect is clearly observable. A quantitative analysis for K$_3$C$_{60}$ yields the vortex pinning force constant that can be hardly determined by other means. Our observation allowed us to explain long-standing microwave anomalies in superconductors \cite{SimonPRB2005,MuranyiPRB2008} and it may lead to pertinent applications in microwave communication techniques.

\section*{Acknowledgements} Stimulating discussions with K\'aroly Holczer are appreciated. Support by the National Research, Development and Innovation Office of Hungary (NKFIH) Grant Nrs. K119442, SNN118012 and 2017-1.2.1-NKP-2017-00001 are acknowledged. This work (PCC and SLB) was supported by the U.S. Department of Energy, Office of Basic Energy Science, Division of Materials Sciences and Engineering. The research was performed at the Ames Laboratory which is operated for the U.S. Department of Energy by Iowa State University under Contract No. DE-AC02-07CH11358.

\section*{Author Contributions}
GC implemented the CC calculations, which were verified by BGM under the supervision of FS. AJ and FM contributed to the microwave impedance measurements. NMN prepared and characterized the single crystal K$_3$C$_{60}$. GK and KK prepared and characterized the fine powder K$_3$C$_{60}$ samples. SLB and PCC prepared and characterized the MgB$_2$ sample. VGK helped the theoretical discussion in the paper. All authors contributed to writing of the manuscript.





\appendix
\newpage
\pagebreak
\clearpage

\title{Supplementary Material for: Giant microwave absorption in fine powders of superconductors}

\author{G. Cs\H{o}sz}
\affiliation{Department of Physics, Budapest University of Technology and Economics and
MTA-BME Lend\"{u}let Spintronics Research Group (PROSPIN), POBox 91, H-1521 Budapest, Hungary}

\author{B. G. M\'{a}rkus}
\affiliation{Department of Physics, Budapest University of Technology and Economics and
MTA-BME Lend\"{u}let Spintronics Research Group (PROSPIN), POBox 91, H-1521 Budapest, Hungary}

\author{A. J\'{a}nossy}
\affiliation{Department of Physics, Budapest University of Technology and Economics and
MTA-BME Lend\"{u}let Spintronics Research Group (PROSPIN), POBox 91, H-1521 Budapest, Hungary}

\author{N. M. Nemes}
\affiliation{GFMC, Unidad Asociada ICMM-CSIC "Laboratorio de Heteroestructuras con Aplicacion en Espintronica", Departamento de Fisica de Materiales Universidad Complutense de Madrid, 28040 Madrid, Spain}

\author{F. Mur\'{a}nyi}
\affiliation{Mettler-Toledo GmbH, Heuwinkelstrasse 3., CH-8606 N\"anikon, Switzerland}

\author{G. Klupp}
\affiliation{Institute for Solid State Physics and Optics, Wigner
Research Centre for Physics, Hungarian Academy of Sciences, P.O. Box
49, H-1525 Budapest, Hungary}

\author{K. Kamar\'{a}s}
\affiliation{Institute for Solid State Physics and Optics, Wigner
Research Centre for Physics, Hungarian Academy of Sciences, P.O. Box
49, H-1525 Budapest, Hungary}

\author{V. G. Kogan}
\affiliation{Ames Laboratory, U.S. Department of Energy and Department of Physics and Astronomy, Iowa State University,
Ames, Iowa 50011, USA}

\author{S. L. Bud'ko}
\affiliation{Ames Laboratory, U.S. Department of Energy and Department of Physics and Astronomy, Iowa State University,
Ames, Iowa 50011, USA}

\author{P. C. Canfield}
\affiliation{Ames Laboratory, U.S. Department of Energy and Department of Physics and Astronomy, Iowa State University,
Ames, Iowa 50011, USA}

\author{F.\ Simon}
\email{f.simon@eik.bme.hu}
\affiliation{Department of Physics, Budapest University of Technology and Economics and
MTA-BME Lend\"{u}let Spintronics Research Group (PROSPIN), POBox 91, H-1521 Budapest, Hungary}

\maketitle

\section{The sum rule and complex conductivity in superconductors}

For usual metals, the sum rule states \cite{KuboJSPS1957}:
\begin{equation}
	\frac{2}{\pi}\int_{0}^{\infty}\sigma_1\left(\omega\right)\dd\omega=\frac{ne^2}{m^{\ast}}=\epsilon_\text{0}\Omega_{\text{pl}}^2,
	\label{eq:sum rule}
\end{equation}
where $n$ is the charge carrier density, $e$ is the elementary charge, $m^{\ast}$ is the effective mass, $\Omega_{\text{pl}}$ is the plasma frequency, and $\epsilon_\text{0}$ is the vacuum dielectric constant. This is satisfied for the AC Drude model, where $\sigma_{1}$ and $\sigma_{2}$ read:
\begin{gather}
	\sigma_{1}=\frac{ne^2\tau}{m^{\ast}}\frac{1}{1+\omega^2\tau^2},\\
	\sigma_{2}=\frac{ne^2\tau}{m^{\ast}}\frac{\omega\tau}{1+\omega^2\tau^2}.
\end{gather}

For superconductors, the Ferrell--Glover--Tinkham (FGT) theory states that the sum rule \cite{FerrelGloverPRB1958,TinkhamFerrellPR1959} is still obeyed, i.e. the spectral weight is only \emph{re-arranged} due to superconductivity but the value of the integral is retained.

In the following, we discuss the relevance of the sum rule in different superconductor regimes. In a type II superconductor, $\lambda\gg (\xi_\text{0},\ell)$,where $\xi_\text{0}$ is the coherence length and $\ell$ is the mean free path, therefore local electrodynamics can be used. For such a case, $\sigma_1(\omega)=0$ below the gap frequency, $\omega_\text{g}=\frac{2\Delta}{\hbar}$ except for the delta function at $\omega=0$. The clean limit is defined as $\xi_0\ll\ell$, which is equivalent to $\omega_\text{g}\gg 1/\tau$, since $\xi_0=\hbar v_{\text{F}}/\pi \Delta$. In this case, all the oscillator strength appears in the $\sigma_1\propto\delta(\omega)$ function as:
\begin{gather}
	\sigma_{1,\text{clean}}=\frac{ne^2}{m^{\ast}}\delta\left(\omega\right)=\frac{\pi}{2\mu_\text{0}\lambda_\text{L}^2}\delta\left(\omega\right),\\
	\sigma_{2,\text{clean}}=\frac{1}{\mu_\text{0}\omega\lambda_\text{L}^2}, 
	\label{eq:sigma_clean}
\end{gather}
where $\lambda_\text{L}$ is the London penetration depth, related to the charge carrier concentration as: $\lambda_\text{L}=\sqrt{\frac{m^{\ast}}{\mu_0 n e^2}}$ and also $\lambda_\text{L}=c/\Omega_{\text{pl}}$ (where $c$ is the speed of light). Note that $\sigma_{1,\text{clean}}$ clearly satisfies the sum rule. 

In the dirty limit, $\xi_0\gg\ell$ (which is the case in K$_\text{3}$C$_\text{60}$, given that $\xi_0\approx 3$ nm and $\ell=1$ nm)\cite{GunnRMP}, we have $\omega_\text{g}\ll 1/\tau$. Then, the oscillator strength in the delta function is reduced by approximately $\omega_\text{g}\tau$ (which is smaller than $1$). Pippard's aproximate expression for the penetration depth, $\lambda$, in the dirty limit
\begin{equation}
	\lambda=\lambda_\text{L}\sqrt{1+\frac{\xi_\text{0}}{\ell}},
	\label{eq:lambda_dirty}
\end{equation}
leads to the same result: 
\begin{gather}
\sigma_{1,\text{dirty}}(\omega=0)= \frac{\lambda_{\text{L}}^2}{\lambda^2}\sigma_{1,\text{clean}}(\omega=0)\approx\\ \nonumber \frac{\ell}{\xi}\sigma_{1,\text{clean}}(\omega=0)=\omega_\text{g}\tau\sigma_{1,\text{clean}}(\omega=0).
\end{gather}

A theoretical consideration of the charge carriers and effective mass in K$_\text{3}$C$_\text{60}$ gave a clean limit result as $\lambda_\text{L}=160$~nm (Ref. \onlinecite{GunnRMP}) which is about a factor $3$ smaller than the experimental value of $\lambda\sim 400$~nm. This means that the delta function strength is reduced by about a factor $10$ as compared to the clean limit case.

In principle, one has two further regimes for type I superconductors, which are however not relevant for the present discussion. When $\xi_\text{0}\ll \ell$, $\sigma_{1}$ and $\sigma_{2}$ have similar form as in the clean limit. When $\xi_\text{0}\gg \ell$, one must use non-local electrodynamics, which also leads to a delta function with a reduced spectral weight. 

\section{Frequency and temperature dependence of the AC conductivity in superconductors}

The description of the AC conductivity was given by Mattis and Bardeen based on BCS theory in 1958.\cite{MattisBardeen} The expressions for $\sigma_1$ and $\sigma_2$ are:
\begin{gather}
	\frac{\sigma_1}{\sigma_\text{n}}=\frac{2}{\hbar\omega}\int_{\Delta}^{\infty}\frac{\left[f\left(\varepsilon\right)-f\left(\varepsilon+\hbar\omega\right)\right]\left(\varepsilon^2+\Delta^2+\hbar\omega\varepsilon\right)}{\left(\varepsilon^2-\Delta^2\right)^{1/2}\left[\left(\varepsilon+\hbar\omega\right)^2-\Delta^2\right]}\dd\varepsilon+ \notag \\\frac{1}{\hbar\omega}\int_{\Delta-\hbar\omega}^{-\Delta}\frac{\left[1-2f\left(\varepsilon+\hbar\omega\right)\right]\left(\varepsilon^2+\Delta^2+\hbar\omega\varepsilon\right)}{\left(\varepsilon^2-\Delta^2\right)^{1/2}\left[\left(\varepsilon+\hbar\omega\right)^2-\Delta^2\right]}\dd\varepsilon, \\ \notag \\
	\frac{\sigma_2}{\sigma_\text{n}}=\frac{1}{\hbar\omega}\int_{\Delta-\hbar\omega,-\Delta}^{\Delta}\frac{\left[1-2f\left(\varepsilon+\hbar\omega\right)\right]\left(\varepsilon^2+\Delta^2+\hbar\omega\varepsilon\right)}{\left(\Delta^2-\varepsilon^2\right)^{1/2}\left[\left(\varepsilon+\hbar\omega\right)^2-\Delta^2\right]}\dd\varepsilon, \text{\label{M_B_2}}
\end{gather}
where $f$ is the usual Fermi-Dirac distribution function, and the lower limit of the integral in (\ref{M_B_2}) for $\hbar\omega>2\Delta$ is $-\Delta$.

\begin{figure}[h!]
	\begin{center}
		\centering
		\includegraphics[width=1\linewidth]{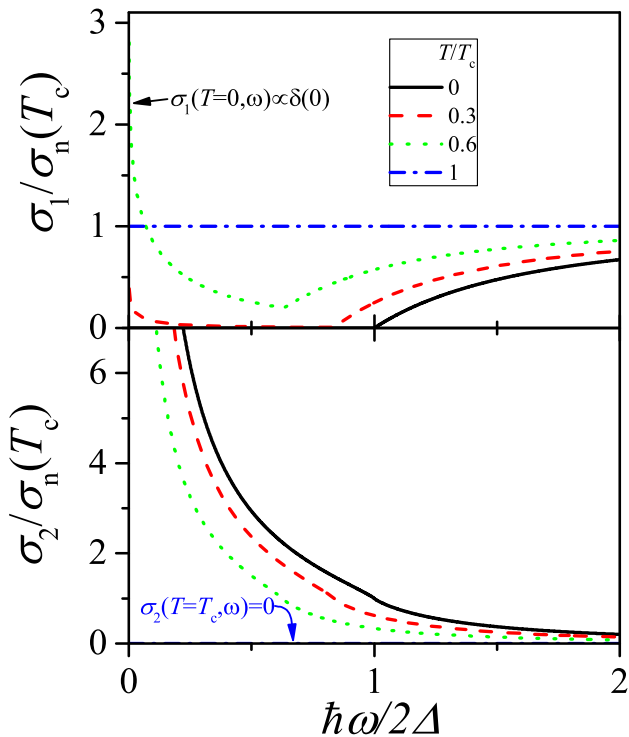}
		\caption{Frequency dependence of the real and imaginary part of the conductivity calculated according to Mattis--Bardeen equations.}
		\label{fig:FigSM_freq_sigma}
	\end{center}
\end{figure}

In Fig. \ref{fig:FigSM_freq_sigma}., we show the real and imaginary parts of the AC conductivity as a function of frequency at some temperatures and $B=0$. Note that $\sigma_2$ vanishes at $T=T_{\text{c}}$ and that $\sigma_1$ becomes a delta function, $\delta(\omega)$, at $T=0$ and $\omega=0$ with a spectral weight which compensates for the oscillator strength which is missing at finite frequencies (this is also known as the sum rule for superconductors\cite{TinkhamBook}). The zero frequency conductivity in the superconducting range therefore equals to roughly: $\sigma_1(\omega)\approx \delta(\omega)\sigma_{\text{n}}\frac{2\Delta}{\hbar}$.

\section{Details of sample preparation and microwave measurement}

Preparation of K$_3$C$_{60}$ proceeds from high purity C$_{60}$ (purity $> 99.9\,\%$, "super gold grade", Hoechsts) and K (Sigma-Aldrich) in stoichiometric amounts. It is heated to $350~^\circ$C in a closed steel capsule for typically 3 days in inert atmosphere followed by grinding of the sample and a final, 1 week-long, heat treatment under the same temperature conditions \cite{FlemingNature}. The sample was repeatedly ground followed by additional heat treatment to enable a homogeneous doping. Characterization was performed using IR spectroscopy, powder X-ray diffractometry, electron spin resonance spectroscopy and SQUID magnetometry.

The starting C$_{60}$ crystal sample was grown with the gradient sublimation technique\cite{DresselhausFullerenes} which results in fullerene single crystals with the size from a few $100$ microns up to $2$ millimeters. The doping then proceeded along the conventional vapor phase doping. The superconducting volume fraction of the K$_3$C$_{60}$ crystal was characterized by SQUID magnetometry and reported in Ref. \onlinecite{NemesKB1999}. The zero-field cooled magnetization reveals \cite{NemesKB1999}, by field exclusion, a superconducting volume fraction close to 100 \%. Note that this method can not reveal core-shell-like arrangements of cores of non-superconducting impurities shielded by superconducting shells. This, however, would not affect the present studies which are also limited to the surface of the K$_3$C$_{60}$ crystal.

In the microwave measurements, we set $\Delta f/f_0=0$ at $T=T_{\text{c}}$ as $\Delta f$ is affected by the quartz tube itself and the amount of helium in the cryostat, i.e. reference measurements do not give a reproducible result. This effect also influences the analysis of the data: an additional fitting parameter, the cavity shift value at $T=T_{\text{c}}$, needs to be introduced. Its value is $+6.7(2)\times 10^{-6}$, i.e. this value is subtracted from the calculated shift values which returns $\Delta f/f_0=0$ at $T=T_{\text{c}}$. We found that this extra parameter does not influence the validity of the numerical analysis of the data.

\section{Additional experimental data}

\begin{figure}[h!]
\begin{center}
  \includegraphics[width=.9\linewidth]{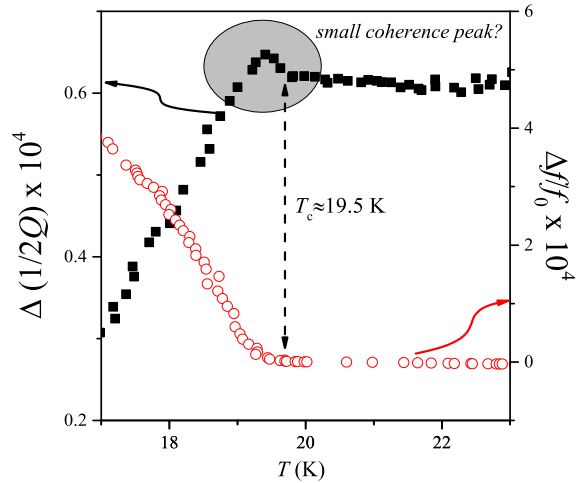}
\caption{Cavity loss and shift as a function of temperature at the Earth's magnetic field. These are the same data shown in the main text, however with a magnified scale. A small peak, which was tentatively assigned to superconducting coherence effects, is observed immediately below $T_{\text{c}}$.}
\label{fig:FigSM_coherencepeak}
\end{center}
\end{figure}

In Fig. \ref{fig:FigSM_coherencepeak}., we show the cavity loss and shift as a function of temperature in zero magnetic field. The data are zoomed to the vicinity of the critical temperature. A small coherence peak is clearly observed immediately below $T_{\text{c}}$. We are aware that this small coherence peak was observed in K$_3$C$_{60}$ and also in Rb$_3$C$_{60}$ ($T_{\text{c}}=28$ K) back in 1994 but its observation remained unpublished \cite{JanossyLegeza}. In principle, a coherence peak is expected in the microwave conductivity as it is predicted by the Mattis--Bardeen theory and similar data were obtained in Ref. \onlinecite{HolczerPRB} for conventional weak-coupled superconductors. However, K$_3$C$_{60}$ is a strong-coupled superconductor \cite{GunnRMP}, therefore the suppression of the conductivity coherence peak is not surprising.

\begin{figure}[h!]
\begin{center}
  \includegraphics[width=.9\linewidth]{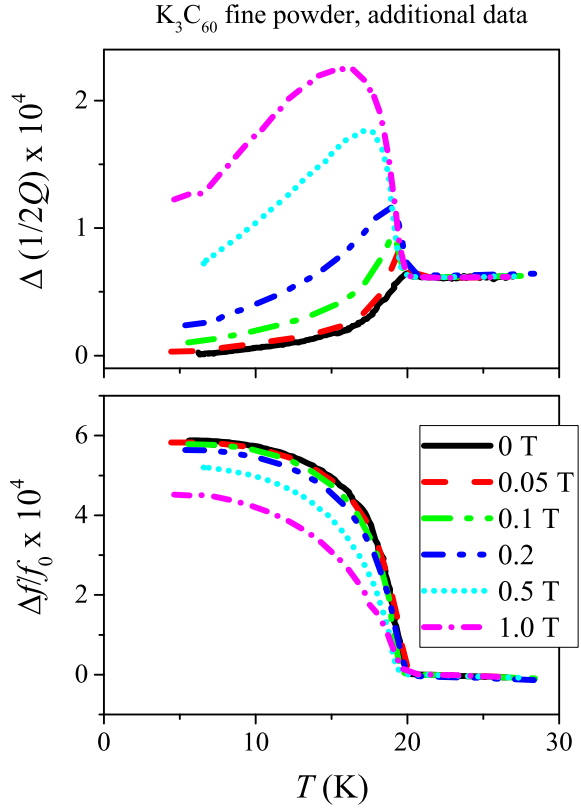}
\caption{Additional experimental data on the fine powder K$_3$C$_{60}$ sample. The data at 0, 0.1, and 1 T are the same as in the main text.}
\label{fig:FIGSM_additional_data}
\end{center}
\end{figure}

In Fig. \ref{fig:FIGSM_additional_data}., we show additional magnetic field dependent cavity loss and shift data for the fine powder K$_3$C$_{60}$ sample. The data at at 0, 0.1, and 1 T are identical to that shown in the main text.

\section{The Coffey--Clem theory}

The Coffey--Clem theory \cite{CC91,CC921,CC922,CC923,CC924,CC93} describes the high frequency electrodynamic response of type-II-superconductors in the mixed state in external static magnetic field. In this calculation the static ($\mathbf{B}_\text{0}$) and the \textit{rf} magnetic field are parallel and the static field is much greater than the amplitude of the \textit{rf} field. This geometry matches our experimental setup. The corresponding \textit{rf} electric field is responsible for the vortex movement and is perpendicular to the static magnetic field. The theory also assumes a uniform vortex density distribution.

To estimate the complex AC conductivity, the two-fluid equation $\mathbf{J}=\mathbf{J_\text{n}}+\mathbf{J_\text{s}}$ is used, where $\mathbf{J_\text{n}}$ is the normal current density, and $\mathbf{J_\text{s}}$ is supercurrent density, which is caused by the vortex dynamics. The current density is characterized by the complex penetration depth, which is defined as follows:
\begin{gather}
\mathbf{J}=\widetilde{\sigma}\mathbf{E}, \\
\widetilde{\sigma}=\frac{\cplxi}{\mu_{\text{0}}\omega\widetilde{\lambda}^2}.
\end{gather}

The quantity $\widetilde{\lambda}$ is determined from the current densities as follows. The two types of current density are obtained from $\mathbf{J_\text{n}}=\sigma_\text{nf}\mathbf{E}$ and the London equations: 

\begin{equation}
\nabla\times\mathbf{J_\text{s}}=-\frac{1}{\mu_{\text{0}}\lambda^2}\left(\mathbf{B}-\mathbf{B}_\text{v}\right),
\end{equation}
where $\mathbf{B}=\mathbf{B}_\text{0}+\mathbf{B}_\text{rf}$, $\mathbf{B}_\text{v}=n\Phi_\text{0}\mathbf{\widehat{B}}_\text{0}$ is the total magnetic field in the vortices, which is the sum of the external DC field ($\mathbf{B}_\text{0}$) and the vortex-motion-induced magnetic field. $\widehat{\mathbf{B}}_\text{0}$ is the direction of the DC magnetic field, $n$ denotes the local area density of vortices. These relationships together with Ampere's law yield:

\begin{equation}
\nabla^2\mathbf{B}=\mu_{\text{0}}\sigma_\text{nf}\dot{\mathbf{B}}+\frac{1}{\lambda^2}\left(\mathbf{B}-\mathbf{B}_\text{v}\right).
\label{eq:difflond}
\end{equation}

The last term can be obtained from the vortex equation of motion:

\begin{gather}
\mu\ddot{\mathbf{u}}+\eta\dot{\mathbf{u}}+\kappa_p\mathbf{u}=\mathbf{J}\times\Phi_\text{0}\mathbf{\widehat{B}_\text{0}},
\label{eq:vortex_eq_motion}
\end{gather}
where $\mathbf{u}$ is the vortex displacement from its equilibrium pinning site, $\mu\left(T\right)=\epsilon_\text{0}\Phi_\text{0}B_\text{c2}\left(T\right)$ is the inertial mass per unit length of vortex, $\eta$ is the viscous drag coefficient in the absence of flux creep and $\kappa_\text{p}$ is the restoring force constant of a pinning potential. All these parameters depend on temperature.

The viscous drag coefficient is directly determined by the other parameters as \cite{Anderson,WuSridhar,PatrickLaurent}:
\begin{gather}
\eta\left(T\right)=\frac{\Phi_\text{0}B_\text{c2}\left(T\right)}{\rho_\text{n}\left(T\right)}.\\
\end{gather}
However, only the maximum value for the pinning force constant can be estimated\cite{Anderson,WuSridhar,PatrickLaurent}. It is obtained for a perfect pinning center, which is a hollow core with radius $\xi$:
\begin{gather}
\kappa_{\text{p,max}}=\frac{B_{\text{c1}} B_{\text{c2}}}{2\mu_0} /\ln\left(\frac{\lambda}{\xi}\right),
\end{gather}
where $\mu_0$ is the vacuum permeability. In a realistic case, $\kappa_{\text{p}}$ can be substantially smaller than $\kappa_{\text{p,max}}$ depending on the quality of the pinning centers.

For a harmonic \textit{rf} excitation, Eq. \eqref{eq:vortex_eq_motion} can be rewritten with the help of the complex dynamic vortex mobility $\widetilde{\mu}_\text{v}=\left(-\cplxi\omega\mu+\eta+\cplxi\kappa_p/\omega\right)^{-1}$ as:
\begin{gather}
\mathbf{u}=-\frac{\widetilde{\mu}_\text{v}}{\cplxi\omega}\mathbf{J}\times\Phi_\text{0}\mathbf{\widehat{B}_\text{0}}.
\label{eq:vortex_eq_motion_rewritten}
\end{gather}
Note the distinction between $\mu$ (inertial mass per unit length of vortex), $\mu_\text{v}$ (complex dynamic vortex mobility), and $\mu_0$ (permeability of the vacuum).

When $\mathbf{J}$ and $\mathbf{\widehat{B}_\text{0}}$ are perpendicular, and the vortex-motion-induced electric field is $\mathbf{E}_\text{v}=\mathbf{B}_\text{v}\times\dot{\mathbf{u}}$ which obeys $\frac{\partial \mathbf{B}_\text{v}}{\partial t}=-\nabla\times\mathbf{E}_\text{v}$:
\begin{equation}
	\mathbf{B}_\text{v}=\mathbf{B}_\text{0}-\nabla\times\left(\mathbf{B}_\text{0}\times\mathbf{u}\right).
	\label{eq:vortex_magnetic_field}
\end{equation}

Considering the vector calculus identities ($\nabla\times\left(\nabla\times\mathbf{B}\right)=\nabla\left(\nabla\mathbf{B}\right)-\nabla^2\mathbf{B}$) and  Faraday's law with harmonic \textit{rf} excitation ($-\nabla\times\mathbf{E}=\dot{\mathbf{B}}=-\cplxi\omega\mathbf{B}$) with the help of equation (\ref{eq:vortex_magnetic_field}), equation (\ref{eq:difflond}) can be rewritten:
\begin{widetext}
\begin{equation}
-\nabla\times\left(\nabla\times\mathbf{B}\right)=-\mu_{\text{0}}\sigma_\text{nf}\nabla\times\mathbf{E}+\frac{1}{\lambda^2}\left(\frac{1}{\cplxi\omega}\nabla\times\mathbf{E}+\nabla\times\left(\mathbf{B}_\text{0}\times\mathbf{u}\right)\right).
\end{equation}
\end{widetext}

The components of this equation can be integrated with respect to the space variable:

\begin{gather}
\mu_{\text{0}}\mathbf{J}=\mu_{\text{0}}\sigma_\text{nf}\mathbf{E}+\frac{1}{\lambda^2}\left(-\frac{1}{\cplxi\omega}\mathbf{E}+\frac{\Phi_\text{0}B_\text{0}\widetilde{\mu}_\text{v}}{\cplxi\omega}\mathbf{J}\right), \label{eq:difflondon_rewrite} \\
\mathbf{J}=\frac{\mu_{\text{0}}\sigma_\text{nf}-\frac{1}{\lambda^2 \cplxi\omega}}{\mu_{\text{0}}-\frac{\Phi_\text{0}B_\text{0}\widetilde{\mu}_\text{v}}{\cplxi\omega\lambda^2}}\mathbf{E}, \\
\widetilde{\lambda}^2=\frac{\lambda^2+(\cplxi/2)\widetilde{\delta}_\text{vc}^2}{1-2 \cplxi\lambda^2/\widetilde{\delta}_\text{nf}^2}.
\end{gather}
The latter quantity is the so-called complex penetration depth. In these equations, $\widetilde{\delta}_\text{nf}$ is the normal fluid skin depth, and $\widetilde{\delta}_\text{vc}^2=\frac{2B_0\Phi_\text{0}\widetilde{\mu}_\text{v}}{\mu_{\text{0}}\omega}$ is the complex effective skin depth. The latter quantity is zero for $B=0$ and it becomes finite in the mixed state when vortex motion is present. Note that for $B=0$, i.e. $\widetilde{\delta}_\text{vc}^2=0$, $\widetilde{\lambda}^2$ returns $\lambda^2$ and $\cplxi \delta_{\text{n}}^2/2$ at $T=0$ and $T=T_{\text{c}}$, respectively.

The temperature and magnetic field dependence of the transport and magnetic parameters are:
\begin{gather}
	B_\text{c2}\left(T\right)=B_\text{c2}\left(0\right)\frac{1-t^2}{1+t^2},\\
	\lambda\left(T,B\right)=\frac{\lambda\left(0,0\right)}{\sqrt{\left(1-b\left(T\right)\right)\left(1-t^4\right)}},\\
	\widetilde{\delta}_{\text{nf}}^2\left(T,B\right)=\delta_{\text{n}}^2/f\left(T,B\right),\\
	f\left(T,B\right)=1-\left(1-t^4\right)\left(1-b\left(T\right)\right),\\
	\kappa_{\text{p}}\left(T\right)=\kappa_{\text{p}}\left(0\right)\left(1-t^2\right)^2,
\end{gather}
where $t=T/T_\text{c}$ and $b(T)=B/B_\text{c2}(T)$ are the reduced temperature and magnetic field, respectively.

\begin{figure}[h!]
\begin{center}
  \includegraphics[width=.9\linewidth]{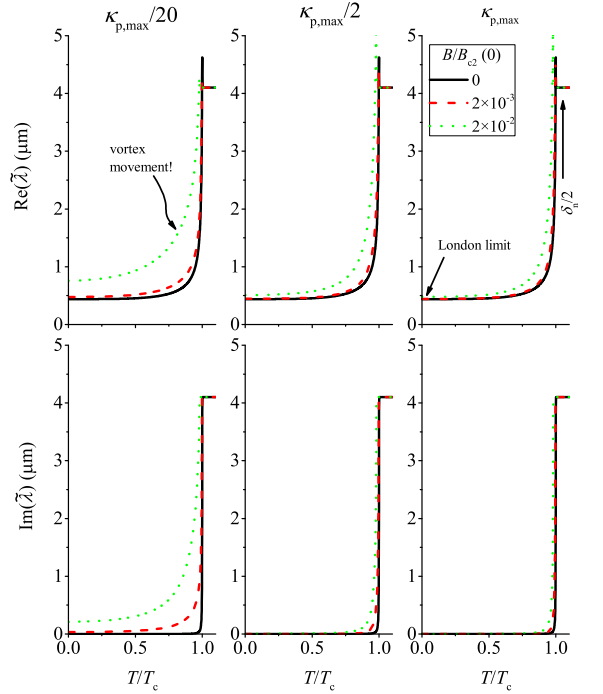}
\caption{Temperature dependence of the real and imaginary parts of $\widetilde{\lambda}$ for various magnetic fields and $\kappa_{\text{p}}$ values. Note that $\text{Re}\left(\widetilde{\lambda}\right)$ returns the London penetration depth at zero magnetic field and also shows the effect of vortex movement for a finite $B$. In the normal state, it returns $\delta_{\text{n}}/2$ as expected.}
\label{fig:FIGSM_lambda_tilde}
\end{center}
\end{figure}

It is informative to show the temperature dependence $\widetilde{\lambda}$, which correspond to the presented $\widetilde{\sigma}$ in the main text. This is shown for various magnetic fields and $\kappa_{\text{p}}$ values in Fig. \ref{fig:FIGSM_lambda_tilde}. In zero magnetic field, $\text{Re}\left(\widetilde{\lambda}\right)$ drops rapidly from $\delta_{\text{n}}/2$ at $T=T_{\text{c}}$ to the London penetration depth toward $T=0$. However, for finite magnetic fields and a smaller $\kappa_{\text{p}}$, $\text{Re}\left(\widetilde{\lambda}\right)$ remains closer to the normal state value down to about half of $T_{\text{c}}$ due to the vortex induced conductivity. For small grain sizes, this allows for a significant microwave field penetration into the grains which leads to a sizeable microwave cavity loss such as we observe herein. 

\section{Additional discussion of the vortex motion in external static and \textit{rf} magnetic field}

\begin{figure}[h!]
	\begin{center}
		\includegraphics[width=.9\linewidth]{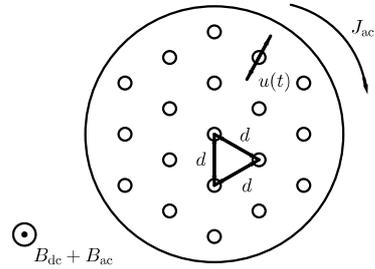}
		\caption{Sphere superconductor in external static and \textit{rf} magnetic field. The magnetic field is parallel to the $z$ axis, and the induced current density is tangential. So the direction of the vortex motion is radial.}
		\label{fig:FigSM_sphere}
	\end{center}
\end{figure}

In the following discussion we assume that the complex conductivity is dominated by the flux flow. So we neglect the effect of vortex pinning and flux creep. In this case the vortex motion is retarded only by the viscous damping. So we can obtain the vortex motion in the following form \cite{TinkhamBook}:
\begin{equation}
	\dot{\mathbf{u}}=\frac{1}{\eta}\mathbf{J}\times\Phi_\text{0}\mathbf{\widehat{B}_\text{0}}.
	\label{eq:vortex_velocity}
\end{equation}
The current density is calculated from Ohm's law in which the electric field is induced by the external \textit{rf} magnetic field:
\begin{equation}
	\nabla\times\mathbf{J}_\text{ac}=-\frac{1}{\rho_\text{f}}\dot{\mathbf{B}}_\text{ac},
	\label{eq:current_density}
\end{equation}
where $\rho_\text{f}$ is the flux-flow resistivity  $\left(\rho_\text{f}=\frac{B\Phi_\text{0}}{\eta}\right)$. For sphere samples the solution of the (\ref{eq:current_density}) equation is the following in cylindrical coordinates:
\begin{equation}
	\mathbf{J}_\text{ac}=-\frac{r}{2\rho_\text{f}}\dot{\mathbf{B}}_\text{ac},
\end{equation}
The directions of the vectors are shown in Fig. \ref{fig:FigSM_sphere}.
Equation (\ref{eq:vortex_velocity}) can be rewritten with this result:
\begin{equation}
	\dot{u}=-\frac{r}{2}\frac{\dot{B}_\text{ac}}{B}.
\end{equation}

This result is the same, if we assume that vortex motion is caused by the change of vortex density. The reason of this change is the small oscillation in the magnetic field (small so that only the size of the primitive cell of the vortex lattice varies). The only relevant length scale is the distance between the vortices, which is proportional to $\left(\frac{\Phi_\text{0}}{B}\right)^{1/2}$.\cite{TinkhamBook} The motion of the vortices can only be radial due to the lattice symmetry. Thus the velocity of the vortices is equal to the time derivative of $r$, where:
\begin{equation}
	r=n\left(\frac{\Phi_\text{0}}{B}\right)^{1/2},
\end{equation}
where $n$ depends only on the vortex lattice structure. The time derivative of $r$ is:
\begin{equation}
	\dot{r}=-\underbrace{n\left(\frac{\Phi_\text{0}}{B}\right)^{1/2}}_{r}\frac{\dot{B}_\text{ac}}{2B}=-\frac{r}{2}\frac{\dot{B}_\text{ac}}{B}.
\end{equation}

\section{Frequency dependence of the conductivity in the presence of vortex motion}
It is assumed that conductivity can be written as a sum of two terms: $\sigma=\sigma_{\text{nf}}+\sigma_{\text{sf}}$, where $\sigma_{\text{nf}}$ and $\sigma_{\text{sf}}$ are the normal-fluid and superconductor contributions, respectively. At low temperatures, the normal fluid term $\left(\lambda^2\ll\delta_{\text{nf}}^2\right)$ can be neglected, thus Eq. \eqref{eq:difflond} can be rewritten as follows:
\begin{equation}
	\mu_{\text{0}}\lambda^2\dot{\mathbf{J}}+\frac{1}{\sigma_\text{v}\left(\omega\right)}{\mathbf{J}}=\mathbf{E}.
\end{equation}
Note, that this equation is very similar to the kinetic equation of electrons in the AC Drude model:	$\frac{m}{ne^2}\dot{\mathbf{J}}+\frac{1}{\sigma_{\text{0}}}{\mathbf{J}}=\mathbf{E}$.

According to the CC theory, the value of $\sigma_{\text{v}}$ can be written as\cite{CC91}:
\begin{equation}
	\sigma_{\text{v}}\left(\omega\right)=\sigma_{\text{f}}\left(1+\frac{\kappa_p}{\cplxi\omega\eta}\right),
\end{equation}
where $\sigma_{\text{f}}=\frac{B_\text{c2}}{B}\sigma_{\text{n}}$ is the flux-flow conductivity (which appears in the Bardeen--Stephen theory, i.e. it is without vortex pinning). 

\begin{figure}[h!]
	\centering
	\includegraphics[width=1\linewidth]{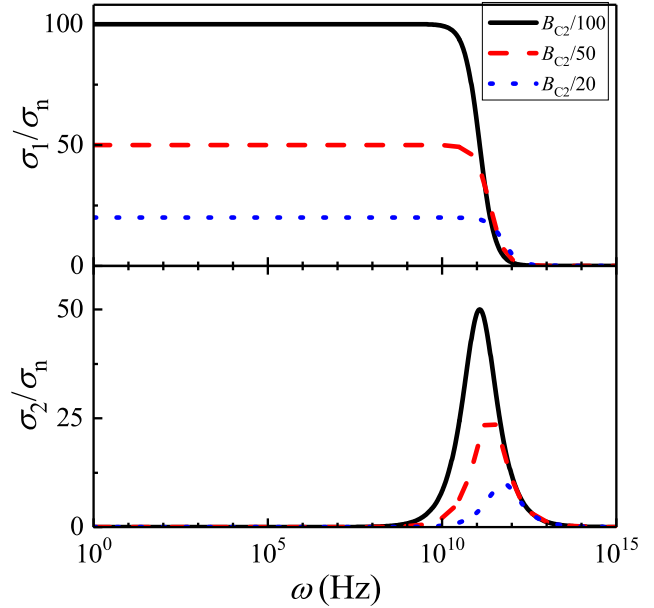}
	\caption{Complex conductivity in different magnetic fields with no pinning centers. Note that when $B\rightarrow0$ we recover the delta function in $\sigma_{1}$. Note, that the scale of the abscissa is logarithmic, as a result the lowering of the edge in the $\sigma_1$ data can be be hardly seen.}
	\label{fig:graphzerokappa}
\end{figure}

In the absence of vortex pinning centers, $\sigma_{\text{v}}=\sigma_{\text{f}}$ i.e. it is the inverse of the Bardeen--Stephen resistivity \cite{BardeenStephen}. In this case $\sigma_{\text{v}}$ does not depend on frequency, so the frequency-dependent conductivity has the same form as the conductivity in the Drude model, shown in  Fig. \ref{fig:graphzerokappa}.
\begin{equation}
	\sigma\left(\omega\right)=\frac{\sigma_{\text{f}}}{1+\cplxi\omega\mu_{\text{0}}\lambda^2\sigma_{\text{f}}}.
	\label{eq:onlyfield}
\end{equation}
Note that \eqref{eq:onlyfield} satisfies the sum rule:
\begin{equation}
	\frac{2}{\pi}\int_{0}^{\infty}\text{Re}\sigma\left(\omega\right)\mathrm{d}\omega=\frac{1}{\pi}\int_{-\infty}^{\infty}\frac{\sigma_\text{f}}{1+\omega^2\left(\mu_{\text{0}}\lambda^2\sigma_{f}\right)^2}\dd\omega=\frac{1}{\mu_\text{0}\lambda^2},
\end{equation} 
where we exploited that $\text{Re}\,\sigma\left(\omega\right)$ is even and the $\frac{1}{1+\omega^2\tau^2}$ function has one simple pole at $\cplxi/\tau$.

The conductivity has a more complicated form according to the CC theory when $\kappa_{\text{p}}\ne0$:
\begin{gather}
\sigma_{1}\left(\omega\right)=\frac{\sigma_{f}\lambda_{\text{C}}^4}{\left(\lambda_{\text{C}}^2\mu_{\text{0}}\omega\lambda^2\sigma_{f}\right)^2+\left(\lambda_{\text{C}}^2+\lambda^2\right)^2}+\frac{\pi}{2}\frac{\delta\left(\omega\right)}{\mu_0 \left(\lambda^2+\lambda_{\text{C}}^2\right)}, \\
	\sigma_{2}\left(\omega\right)=-\frac{\lambda_{\text{C}}^4\mu_{\text{0}}^2\omega^2\lambda^2\sigma_{f}^2+\lambda^2+\lambda_{\text{C}}^2}{\mu_{\text{0}}\omega\left(\left(\lambda_{\text{C}}^2\mu_{\text{0}}\omega\lambda^2\sigma_{f}\right)^2+\left(\lambda_{\text{C}}^2+\lambda^2\right)^2\right)}
\end{gather}
where $\lambda_\text{C}^2=B\Phi_\text{0}/\mu_{\text{0}}\kappa_{\text{p}}$ is the so-called Campbell penetration depth\cite{Campbell1,Campbell2}.

\begin{figure}[h!]
	\centering
	\includegraphics[width=1\linewidth]{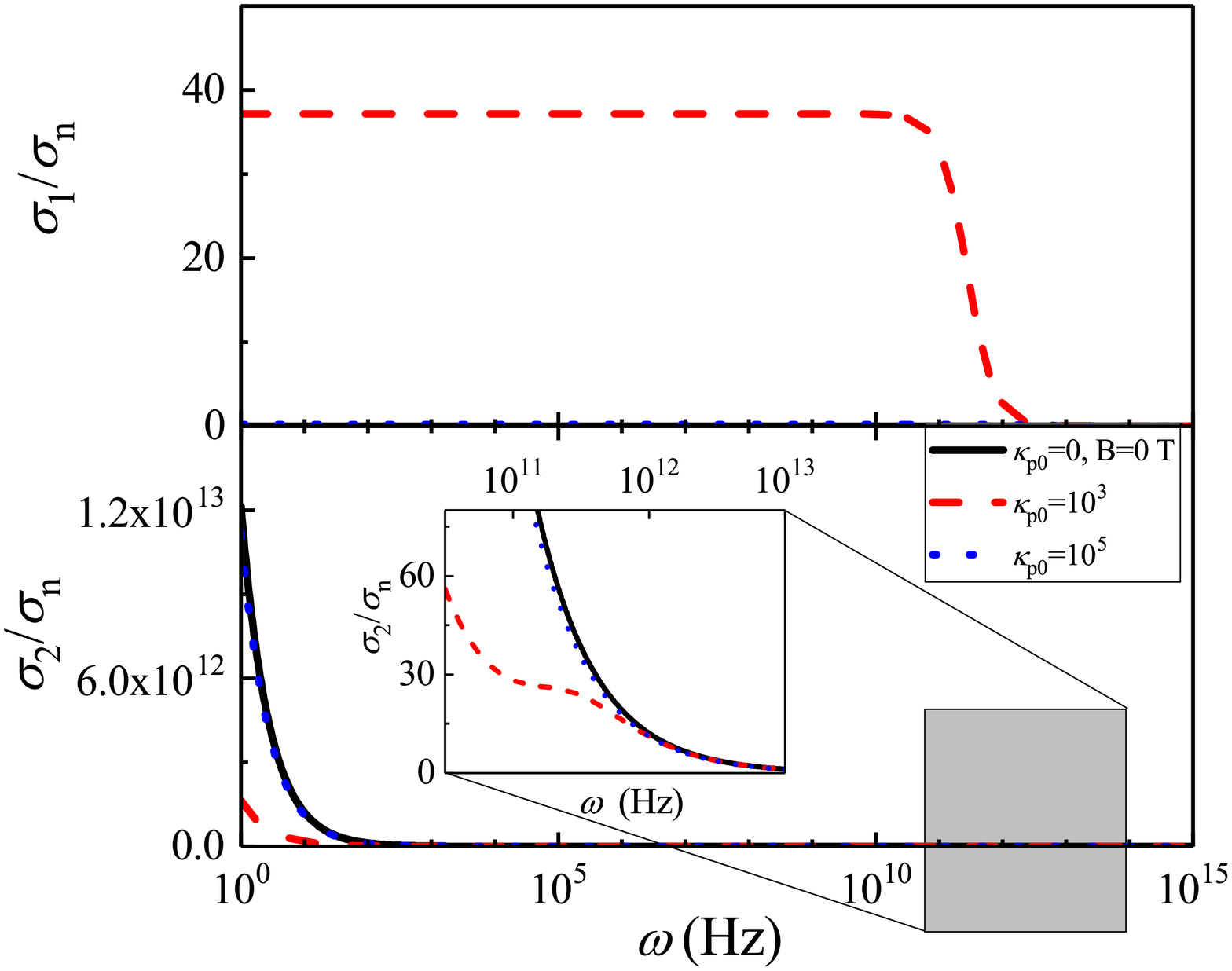}
	\caption{Complex conductivity in finite magnetic field with different pinning forces; a new term, proportional to $1/\omega$, appears in $\sigma_{2}$. For the $\sigma_1$ data, the curves with large $\kappa_{\text{p}}$ and $B=0$ T values remain zero except for a delta function. Note that for a large pinning force, $\sigma_{1}$ and $\sigma_{2}$ have the same form as the conductivity in zero applied magnetic field. The scale of the abscissa is logarithmic. A zoom-in is shown for the grey shaded box in the insert.}
	\label{fig:graphfinitekappap0}
\end{figure}

The CC theory provides the first term only in $\sigma_{1}$. The presence of the second term with the delta function was inserted in order to satisfy the sum rule and to maintain the Kramers--Kronig relation between $\sigma_{1}$ and $\sigma_{2}$. The second terms in $\sigma_{1,2}$ are due to the strongly pinned vortices, i.e. these appear as those in superconductors in the Meissner state. The spectral weight is distributed between the two terms in $\sigma_{1}$: as the pinning force is increased, the delta function dominates and eventually it returns the case of $B=0$. In the other extreme, when $\kappa_{\text{p}}$ tends to zero, the Bardeen--Stephen result is recovered. The full result is shown in Fig. \ref{fig:graphfinitekappap0}. for a few $\kappa_{\text{p}}$ values.

\section{Microwave cavity perturbation in the skin limit}

\begin{figure}[htp]
	\begin{center}
		\centering
		\includegraphics[width=1\linewidth]{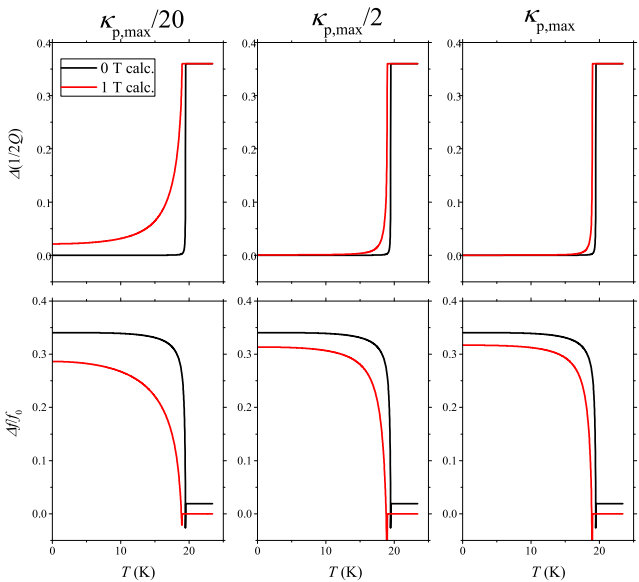}
		\caption{Calculated cavity loss and shift calculated according to Eq. \eqref{eq:SM_skinlimit} using the $\sigma_1$ and $\sigma_2$ data shown in the main text at $f=10\,\text{GHz}$. We used $\nu=1$ in the calculation and assumed that the sample is in the node of the electric field inside a microwave cavity.}
		\label{fig:FIGSM_skin_limit_different_kappa}
	\end{center}
\end{figure}

As discussed in the main text, when the penetration of microwaves is limited to the skin depth only, the microwave cavity loss and shift is expressed by the complex penetration depth $\widetilde{\lambda}$ as:

\begin{gather}
	\frac{\Delta f}{f_\text{0}}-\cplxi\Delta\left(\frac{1}{2Q}\right)=-\cplxi \nu \mu_0 \omega \sqrt{-\widetilde{\lambda}^2}, 
\label{eq:SM_skinlimit}
\end{gather}
where $\nu$ is called resonator constant \cite{HolczerPRB}. 

Fig. \ref{fig:FIGSM_skin_limit_different_kappa}. shows the cavity loss calculated according to Eq. \eqref{eq:SM_skinlimit} for various values of $\kappa_{\text{p}}$ (the same $\sigma_1$ and $\sigma_2$ data as in the main text). Although $\sigma_1$ and $\sigma_2$ changes significantly for the various $\kappa_{\text{p}}$ values, remarkably, in this limit the cavity loss and shift show little sensitivity to $\kappa_{\text{p}}$. As a result, surface impedance studies cannot determine the value of the pinning force constant with certainty.
 
\begin{figure}[h!]
	\begin{center}
		\centering
		\includegraphics[width=1\linewidth]{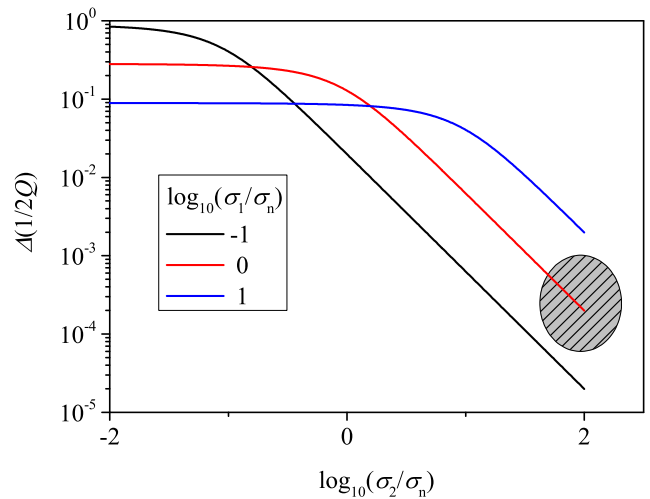}
		\caption{Calculated cavity loss vs normalized $\sigma_\text{2}$ in logarithmic scale for different $\sigma_\text{1}$ values with the geometric factor, $\nu=1$. The shaded area shows the experimentally relevant range of parameters.}
		\label{fig:FIGSM_cavloss_vs_sigma2}
	\end{center}
\end{figure}

Fig. \ref{fig:FIGSM_cavloss_vs_sigma2}. depicts the sensitivity of the cavity loss as a function of $\sigma_2/\sigma_{\text{n}}$. Well below $T_{\text{c}}$, the latter quantity is about $100$. At the same time, $\sigma_1/\sigma_{\text{n}}$ remains around unity if the vortex motion is significant. However, in the skin limit regime, the cavity loss drops by about $2$ orders of magnitude due to the significant $\sigma_2$, which fully prevents any meaningful measurement of $\sigma_1$ in surface impedance studies.

\section{Microwave cavity perturbation in the penetration limit}

\begin{figure}[h!]
	\begin{center}
		\includegraphics[width=1\linewidth]{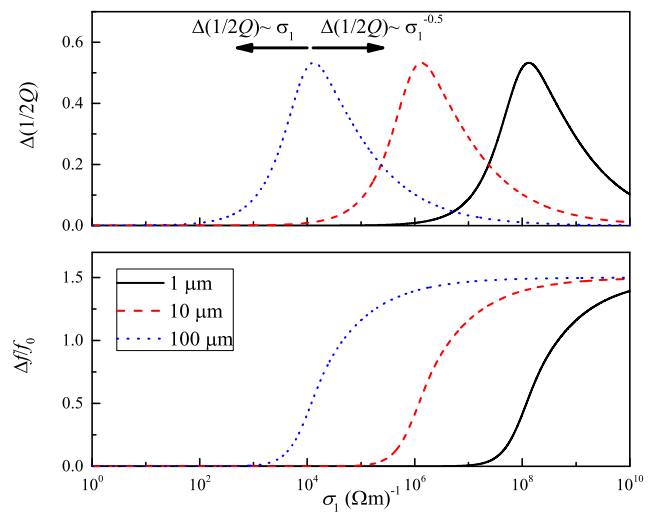}
		\caption{Calculated cavity loss and shift vs. $\sigma_\text{1}$ when $\sigma_2=0$ for different sample sizes with Eq. \eqref{eq:SMpenetr_lim} on a semilogarithmic plot using the geometrical constant $\gamma=1$.}
		\label{fig:FIGSM_loss_diff_sample_size}
	\end{center}
\end{figure}

We mentioned in the main text that for a small particle with diameter $a$ the following relation holds between the cavity parameters and the conductivity \cite{LandauBook}:
\begin{gather}
	\frac{\Delta f}{f_\text{0}}-\cplxi\Delta\left(\frac{1}{2Q}\right)=- \gamma \widetilde{\alpha}, \label{eq:SMpenetr_lim_loss} \\
	\widetilde{\alpha}=-\frac{3}{2}\left(1-\frac{3}{a^2 \widetilde{k}^2}+\frac{3}{a k}\cot(a \widetilde{k}) \right),
\label{eq:SMpenetr_lim}   
\end{gather}
where $\widetilde{k}=\widetilde{n}\frac{\omega}{c}$ is the complex wavenumber with $\widetilde{n}=\sqrt{\frac{\cplxi\widehat{\sigma}}{\epsilon_\text{0}\omega}}$ being the complex index of refraction. Eq. \eqref{eq:SMpenetr_lim} returns $\frac{\Delta f}{f_\text{0}}=0$ and $\Delta\left(\frac{1}{2Q}\right)\propto \sigma$ for small values of $\sigma$ as expected \cite{,Klein1993,MaedaPRL}. In the other extreme, it returns $\frac{\Delta f}{f_\text{0}}=-\gamma \frac{3}{2}$ and $\Delta\left(\frac{1}{2Q}\right)\propto 1/\sqrt{\sigma}$ for large values of $\sigma$, which is also the expected result \cite{,Klein1993,MaedaPRL}.

Fig. \ref{fig:FIGSM_loss_diff_sample_size}. shows the calculated cavity loss and cavity shift as a function of $\sigma_1$ with $\sigma_2=0$ for various particle sizes as calculated with Eq. \eqref{eq:SMpenetr_lim}. This calculation demonstrates that the character of the loss vs $\sigma_1$ changes depending on the value of $\sigma_1$ from a $\Delta\left(\frac{1}{2Q}\right)\propto \sigma_1$ to a $\Delta\left(\frac{1}{2Q}\right)\propto 1/\sqrt{\sigma_1}$, where the characteristic value of the crossover is also particle size dependent. The crossover is accompanied by a change in the cavity shift.

\begin{figure}[h!]
	\begin{center}
	 \includegraphics[width=1\linewidth]{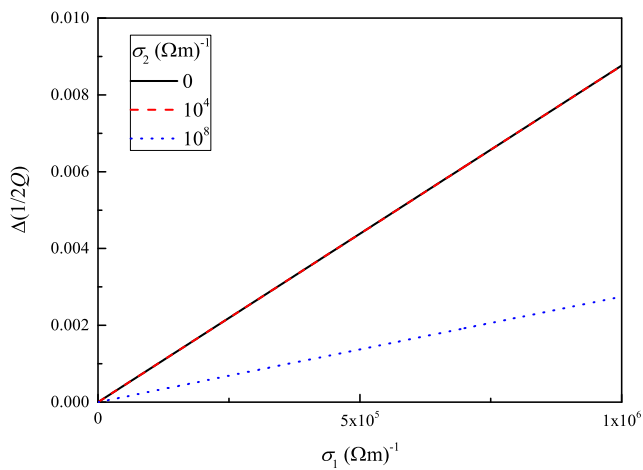}
		\caption{Calculated cavity loss vs. $\sigma_\text{1}$ for different $\sigma_\text{2}$ values. The sample size is $1$ $\mu$m. Cavity loss is in arbitrary units, i.e. multiplied by an arbitrary $\gamma$. Note that the loss becomes less sensitive to $\sigma_\text{1}$ for larger $\sigma_\text{2}$ values (dotted line) but the solid and dashed lines overlap.}
		\label{fig:FIGSM_loss_diff_sigma2}
	\end{center}
\end{figure}

Fig. \ref{fig:FIGSM_loss_diff_sigma2}. demonstrates that the cavity loss decreases for a larger value of $\sigma_2$. This is the reason why in the superconducting state, where $\sigma_2$ is finite, it is more difficult to detect the effect of the vortex motion induced finite $\sigma_1$ on the cavity loss. Nevertheless, the drop in the cavity loss even for a large $\sigma_2=10^8$ $1/\Omega$m is about a factor $3$ as compared to the $2-3$ orders of magnitude cavity loss drop for the same $\sigma_2$ for a single crystal sample. The latter situation is shown in Fig. \ref{fig:FIGSM_cavloss_vs_sigma2}.


\end{document}